\documentclass[useAMS,usenatbib]{mn2e}
\usepackage[dvips]{epsfig}
\def \be{\begin{equation}}
\def \ee{\end{equation}}
\def \bdm{\begin{eqnarray}}
\def \edm{\end{eqnarray}}

\title[Spurious contribution to CR scattering calculations]{Spurious contribution to CR scattering calculations}
\author[A. Shalchi, H. Yan and A. Lazarian]{A. Shalchi$^{1,2}$, H. Yan$^{3}$ and A. Lazarian$^{3}$ \\
$^{1}$Bartol Research Institute, University of Delaware, Newark, Delaware 19716, USA\\
$^{2}$now at: Institut f\"ur Theoretische Physik, Lehrstuhl IV: Weltraum- und Astrophysik, Ruhr-Universit\"at Bochum, D-44780 Bochum, Germany\\
$^{3}$Astronomy Department, University of Wisconsin, Madison, Wisconsin 53706, USA}

\begin{document}

\date{Accepted / Received}

\maketitle

\label{firstpage}

\begin{abstract}
The quasilinear theory for cosmic ray propagation is a well known and widely accepted theory.
In this paper, we discuss the different contributions to the pitch-angle Fokker-Planck coefficient
from large and small scales for slab geometry using the damping model of dynamical turbulence.
These examinations will give us a hint on the limitation range where quasilinear approximation
is a good approximation.
\end{abstract}


\section{Introduction}

The propagation of cosmic rays (CRs) is affected by their interaction
with a magnetic field. This field is turbulent and therefore, the resonant
interaction of cosmic rays with MHD turbulence has been discussed
by many authors as the principal mechanism to scatter and isotropize
cosmic rays (Schlickeiser 2002, Yan \& Lazarian 2002). Although cosmic ray diffusion can
happen while cosmic rays follow wandering magnetic fields (Jokipii
1966), the acceleration of cosmic rays requires efficient scattering.
For instance, scattering of cosmic rays back into the shock is a
vital component of the first order Fermi acceleration (see Longair
1997).
\\[0.5cm]
One important interaction between CRs and MHD turbulence is gyroresonance
scattering. While the propagation of most moderate energy CRs don't sense
slower time variations of turbulence,  turbulence is dynamical for low energy
CRs or when the parallel speed of CRs becomes comparable with Alfv\'en speed
$v_\parallel\sim V_A$. Dynamical MHD turbulence leads to resonance broadening so
that a Breit-Wigner-type function instead of a $\delta$ function should be used
(Schlickeiser \& Achatz 1993, Bieber et al. 1994). In this case CRs interact
with a whole range of turbulence from large to small scales.
\\[0.5cm]
Scattering of CRs cannot happen by the magnetic wave which frequency in the 
particle frame is much lower than the Larmor frequency. Indeed if magnetic field 
changes much slower than the particle gyrorates, the quantity $v_{\perp}^2/B$ 
is preserved (Landau \& Lifshitz 1957). As the result while the pitch-angle of the 
particle changes as the wave passes by the total change of the angle is zero.

In quasilinear theory (QLT) the assumption of unperturbed orbit results in 
non-conservation of the adiabatic invariant $\rho=mv_{\perp}^{2}/2B_{0}$. Whereas
small scale contribution corresponds to the sharp resonance in magnetostatic limit,
the contribution from slow large scale can be overestimated by QLT since the adiabatic
invariant ought to be conserved when the electromagnetic field varies on a time scale
longer than the gyroperiods of CRs $\Omega^{-1}$ (Chandran 2000, 
Yan \& Lazarian 2003).
\\[0.5cm]
In QLT the turbulent field is presented in terms of Fourier modes. It is clear that the 
slow (compared to the particle Larmor frequency) components cannot scatter cosmic 
rays as the consequence of the preservation of the adiabatic invariant. However, this 
effect is not treated in the conventional QLT, which has been widely used to obtained a 
lot of results mostly within the framework of the slab model of turbulence. How reliable 
are those results? We attempt to answer this question within this paper.

Therefore, we will discuss the different contributions from large and small scales for the 
slab model of turbulence. We will give the regimes where the interaction is dominated by large
and small scale contributions. This will give us a hint on the limitation
range where quasilinear approximation is a good approximation.
Because QLT is considered as a standard tool for the calculation of diffusion
coefficients (e. g. Bieber et al. 1994, Chandran 2000, Schlickeiser 2002, Yan \& Lazarian 2003)
it is of great significance to explore the validity of the quasilinear approximation.
\\[0.5cm]
Within quasilinear theory the parallel mean free path $\lambda_{\parallel}$
results from the pitch-angle-cosine ($\mu =p_{\parallel }/p$) average of the inverse of 
the pitch-angle Fokker-Planck coefficient $D_{\mu \mu }$ as (Jokipii 1966, Hasselmann \& 
Wibberenz 1968, Earl 1974)
\be
\lambda_{\parallel}={3 v \over 8} \int_{-1}^{+1} d \mu \; {(1-\mu ^2)^2 \over D_{\mu \mu}(\mu )}.
\label{s0e1}
\ee
The pitch-angle Fokker-Planck coefficient is calculated from the ensemble-averaged first-order
corrections to the particle orbits in the weakly turbulent magnetic field (Hall \& Sturrock 1968)
\be
D_{\mu \mu }(\mu )=Re \int _0^\infty d\xi <\dot {\mu }(t)\dot {\mu}^*(t+\xi )>
\label{s0e2}
\ee
and depends on the nature and statistical properties of the electromagnetic turbulence and the
turbulence-carrying background medium.
\\[0.5cm]
In Sect.2 we derive general expressions for the pitch-angle Fokker-Planck coefficient $D_{\mu \mu }$. 
In Sect. 3 we compare the small scale and large scale contribution to the Fokker-Planck coefficient 
with each other. This results can be used to calculate the cosmic ray parallel mean free path 
using Eq. (\ref{s0e1}) what is done in Sect. 4.
\section{The pitch-angle Fokker-Planck coefficient $D_{\mu \mu}$}
The key input to the calculation of Fokker-Planck coefficients is the correlation tensor
$P_{ij}$. It is common to assume the same temporal behaviour for all tensor components
\be
P_{ij} (\vec{k},t) = P_{ij} (\vec{k},0) \cdot \Gamma (\vec{k},t)
\label{s1e1}
\ee
where $\Gamma$ is the dynamical correlation function. In the past several models for $\Gamma$
were discussed (Schlickeiser 2002). In the current paper we use the damping model of
dynamical turbulence (DT-model, Bieber et al. 1994) for slab geometry:
\be
\Gamma (\vec{k},t) = \Gamma (k_{\parallel},t) = e^{-\alpha \mid k_{\parallel} \mid v_{A} t}.
\label{s1e2}
\ee
Here $v_A$ is the Alfv\'en speed and $\alpha$ is a parameter which allows us to adjust
the strength of dynamical effects. $\alpha=0$ corresponds to the magnetostatic limit
whereas $\alpha=1$ corresponds to strongly dynamical turbulence.

A further important input to our calculations is the wave spectrum. For the following examinations 
we use a simple power spectrum without energy- and dissipation-range:
\bdm
g(k) = \left\{
\begin{array}{ccc}
g_{0} k^{-s} & for & k \geq k_{min} \\
0 & for & k \leq k_{min}
\end{array}
\right.
\label{s1e3}
\edm
with
\be
g_0 = {s-1 \over 8 \pi} \delta B^2 k_{min}^{s-1}.
\label{s1e3a}
\ee
Here and in the following discussions $\delta B$ is the total strength of the turbulent magnetic field and
$B_0$ is the strength of the magnetic background field (mean field).
The parameter $s$ in Eq. (\ref{s1e3}) is the inertial-range spectral index.
For a Kolmogorov-spectrum we have $s=5/3$ which was considered by Teufel \& Schlickeiser 2002 and 2003.
Another choice would be $s=3/2$ (see Cho \& Lazarian 2002).
Assuming this power spectrum, slab geometry and the DT-model the pitch-angle Fokker-Planck coefficient
can be written as (see Teufel \& Schlickeiser 2002)
\bdm
& & D_{\mu \mu} = {(s-1) \Omega^2 (1 - \mu^2) \over 4 k_{min} \alpha v_A} \left( {\delta B \over B_0} \right)^2 \cdot k_{min}^{s} \nonumber\\
&\times&\int_{k_{min}}^{\infty} d k \;
\left[ {k^{-s-1} \over 1+ \left( {k v \mu - \Omega \over \alpha v_A k} \right)^2} + {k^{-s-1} \over 1+ \left( {k v \mu + \Omega \over \alpha v_A k} \right)^2} \right].
\label{s1e4}
\edm
Now we split this integral at the wavenumber $k=\epsilon k_{res}$:
\bdm
& & D_{\mu \mu}^{tot} = {(s-1) \Omega^2 (1 - \mu^2) \over 4 k_{min} \alpha v_A} \left( {\delta B \over B_0} \right)^2 k_{min}^{s} \nonumber\\
& \times & \left\{ \int_{k_{min}}^{\epsilon k_{res}} d k \;
\left[ {k^{-s-1} \over 1+ \left( {k v \mu - \Omega \over \alpha v_A k} \right)^2} + {k^{-s-1} \over 1+ \left( {k v \mu + \Omega \over \alpha v_A k} \right)^2} \right] \right. \nonumber\\
& + & \left. \int_{\epsilon k_{res}}^{\infty} d k \;
\left[ {k^{-s-1} \over 1+ \left( {k v \mu - \Omega \over \alpha v_A k} \right)^2} + {k^{-s-1} \over 1+ \left( {k v \mu + \Omega \over \alpha v_A k} \right)^2} \right] \right\}.
\label{s1e5}
\edm
The resonant wavenumber $k_{res}$ is defined through (see Yan \& Lazarian 2002)
\be
k_{res} = {\Omega \over v \mu} = {1 \over \mu R_{L}}
\label{s1e6}
\ee
where $R_L$ is the gyroradius. In the first integral we have $k^{-1} \geq \epsilon^{-1} \mu R_L$ and the inverse wavenumber
is larger than several gyroradii if $\epsilon \ll 1$. We call this case the large scales. In the second integral we have
$k^{-1} \leq \epsilon^{-1} \mu R_L$ and the invers wavenumber is smaller than several gyroradii. We call this case the
small scales. We always have
\be
{k_{min} \over k_{res}} = \mu R
\label{s1e7}
\ee
where we used the dimensionless rigidity $R=R_{L} k_{min}$. In the current paper we restrict 
our analysis to medium rigidities ($R \ll \epsilon \ll 1, v \gg \alpha v_{A}$) and we have therefore
\be
k_{min} \ll \epsilon \cdot k_{res}.
\label{s1e8}
\ee
In this and the following equations we use $\mu = \mid \mu \mid$, $0 \leq \alpha \leq 1$ and $1<s<2$. With
\be
D_{\mu\mu}^{0} := {(s-1) \Omega^2 (1 - \mu^2) \over 4 k_{min} \alpha v_A} \left( {\delta B \over B_0} \right)^2
\label{s1e9}
\ee
we find
\be
{D_{\mu\mu}^{tot} \over D_{\mu\mu}^{0}} = {D_{\mu\mu}^{LS} \over D_{\mu\mu}^{0}} + {D_{\mu\mu}^{SS} \over D_{\mu\mu}^{0}}
\label{s1e10}
\ee
where we used the large scale
\bdm
&{D_{\mu\mu}^{LS} \over D_{\mu\mu}^{0}} = k_{min}^{s} \int_{k_{min}}^{\epsilon k_{res}} d k \;
\left[ {k^{-s-1} \over 1 + \left( {k v \mu - \Omega \over \alpha v_A k} \right)^2}
+ {k^{-s-1} \over 1+ \left( {k v \mu + \Omega \over \alpha v_A k} \right)^2} \right]
\label{s1e11}
\edm
and the small scale Fokker-Planck coefficient
\bdm
&{D_{\mu\mu}^{SS} \over D_{\mu\mu}^{0}} = k_{min}^{s} \int_{\epsilon k_{res}}^{\infty} d k \;
\left[ {k^{-s-1} \over 1 + \left( {k v \mu - \Omega \over \alpha v_A k} \right)^2}
+ {k^{-s-1} \over 1+ \left( {k v \mu + \Omega \over \alpha v_A k} \right)^2} \right].
\label{s1e12}
\edm
Now we split the large scale Fokker-Planck coefficient into two parts
\bdm
&{D_{\mu\mu}^{LS} \over D_{\mu\mu}^{0}} = k_{min}^{s} \int_{k_{min}}^{\infty} d k \;
\left[ {k^{-s-1} \over 1 + \left( {k v \mu - \Omega \over \alpha v_A k} \right)^2} + {k^{-s-1} \over 1+ \left( {k v \mu + \Omega \over \alpha v_A k} \right)^2} \right] \nonumber\\
& - k_{min}^{s} \int_{\epsilon k_{res}}^{\infty} d k \;
\left[ {k^{-s-1} \over 1+ \left( {k v \mu - \Omega \over \alpha v_A k} \right)^2} + {k^{-s-1} \over 1+ \left( {k v \mu + \Omega \over \alpha v_A k} \right)^2} \right]
\label{s1e13}
\edm
and use the transformation $x=k_{min} / k$ in the first and $x=\epsilon k_{res} / k$ in the second
integral to obtain
\be
{D_{\mu\mu}^{LS} \over D_{\mu\mu}^{0}} = A - \left( {\mu R \over \epsilon} \right)^{s} \cdot B
\label{s1e14}
\ee
where we defined the both integrals:
\bdm
A & = & \int_{0}^{1} d x \; x^{s-1} \left[ {1 \over 1 + a^2 / R^2 (R \mu + x)^2} \right. \nonumber\\
& + & \left. {1 \over 1 + a^2 / R^2 (R \mu - x)^2} \right], \nonumber\\
B & = & \int_{0}^{1} d x \; x^{s-1} \left[ {1 \over 1 + a^2 \mu^2 / \epsilon^2 (\epsilon + x)^2} \right. \nonumber\\
& + & \left. {1 \over 1 + a^2 \mu^2 / \epsilon^2 (\epsilon - x)^2} \right]
\label{s1e15}
\edm
with
\be
a = {v \over \alpha v_A} \gg 1.
\label{s1e16}
\ee
We can also use the transformation $x=\epsilon k_{res} / k$ to simplify the small scale Fokker-Planck coefficient
\be
{D_{\mu\mu}^{SS} \over D_{\mu\mu}^{0}} = \left( {\mu R \over \epsilon} \right)^{s} \cdot B.
\label{s1e17}
\ee
The total Fokker-Planck coefficient is equal to
\be
{D_{\mu\mu}^{tot} \over D_{\mu\mu}^{0}} = A.
\label{s1e18}
\ee
\subsection{Analytical results}
As demonstrated in Teufel \& Schlickeiser 2002 the both integrals (\ref{s1e15}) can be calculated 
approximatelly for different cases. For medium rigidities we have $\mu R \ll \epsilon \ll 1$ and $a/R \gg 1$
and we find:
\bdm
A (a \mu \ll 1) & \approx & {\pi \over \sin ({\pi s \over 2})} {R^s \over a^s} \nonumber\\
A (1 \ll a \mu) & \approx & \pi {R^s \over a} \mu^{s-1} \nonumber\\
B (a \mu \ll \epsilon \ll 1) & \approx & {2 \over s} \nonumber\\
B (\epsilon \ll a \mu \ll 1) & \approx & {\pi \over \sin({\pi s \over 2})} \left( {\epsilon \over a \mu} \right)^s - { 2 \over 2 - s} \left( {\epsilon \over a \mu} \right)^2 \nonumber\\
B (\epsilon \ll 1 \ll a \mu) & \approx & \pi { \epsilon^s \over a \mu} - {2 \over 2 - s} \left( {\epsilon \over a \mu} \right)^2
\label{s1e19}
\edm
With these approximations we obtain the following expressions for the total, small scale and
large scale Fokker-Planck coefficient:
\bdm
D_{\mu\mu}^{tot} (a \mu \ll 1) & \approx & {\pi (s-1) \over 4 \sin (\pi s / 2)} v k_{min} {\delta B^2 \over B_0^2}
{R^{s-2} \over a^{s-1}} (1-\mu^2) \nonumber\\
D_{\mu\mu}^{tot} (1 \ll a \mu) & \approx & {\pi (s-1) \over 4} v k_{min} {\delta B^2 \over B_0^2}
R^{s-2} \mu^{s-1} (1-\mu^2) \nonumber\\
D_{\mu\mu}^{SS} (a \mu \ll \epsilon \ll 1) & \approx & {(s-1) \over 2 s} v k_{min} {\delta B^2 \over B_0^2} \nonumber\\
& \times & R^{s-2} \left({a \mu \over \epsilon}\right)^{s} (1-\mu^2) \nonumber\\
D_{\mu\mu}^{SS} (\epsilon \ll a \mu \ll 1) & \approx & {\pi (s-1) \over 4 \sin (\pi s / 2)} v k_{min} {\delta B^2 \over B_0^2}
{R^{s-2} \over  a^{s-1}} (1-\mu^2) \nonumber\\
D_{\mu\mu}^{SS} (\epsilon \ll 1 \ll a \mu) & \approx & {\pi (s-1) \over 4} v k_{min} {\delta B^2 \over B_0^2}
R^{s-2} \mu^{s-1} (1-\mu^2) \nonumber\\
D_{\mu\mu}^{LS} (a \mu \ll \epsilon) & \approx & {\pi (s-1) \over 4 \sin (\pi s / 2)} v k_{min} {\delta B^2 \over B_0^2}
{R^{s-2} \over a^{s-1}} (1-\mu^2) \nonumber\\
D_{\mu\mu}^{LS} (\epsilon \ll a \mu) & \approx & {(s-1) \over 2 (2-s)} v k_{min} {\delta B^2 \over B_0^2} \nonumber\\
& \times & \left( {\epsilon \over \mu R} \right)^{2-s} {1 \over a} (1-\mu^2)
\label{s1e20}
\edm
The magnetostatic result ($\alpha=0$) for the total pitch-angle Fokker-Planck coefficient is
\be
D_{\mu\mu}^{MS} = {\pi (s-1) \over 4} v k_{min} {\delta B^2 \over B_0^2} R^{s-2} \mu^{s-1} (1-\mu^2)
\label{s1e21}
\ee
for all values of $\mu$. 
\subsection{Numerical results}
\begin{figure}
\begin{center}
\epsfig{file=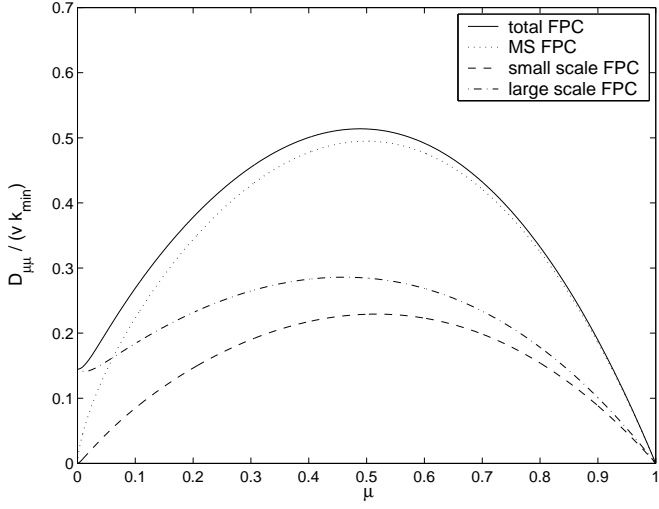, width=250pt}
\end{center}
\caption{Numerical results for the different Fokker-Planck coefficients for protons for $\epsilon=1$. The solid line shows
the total Fokker-Planck coefficient in comparison with the magnetostatic results (dotted line), the small
scale Fokker-Planck coefficient (dashed line) and the large scale Fokker-Planck coefficient 
(dash-dotted line).}
\label{madisonf1}
\end{figure}

\begin{figure}
\begin{center}
\epsfig{file=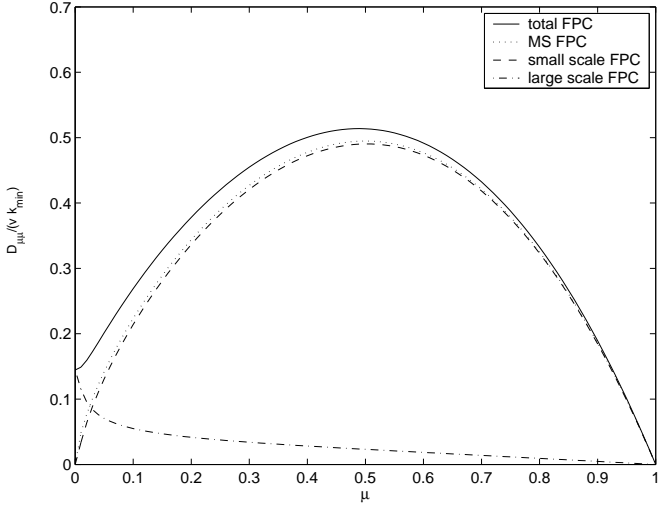, width=250pt}
\end{center}
\caption{Numerical results for the different Fokker-Planck coefficients for protons for $\epsilon=1/3$. The solid line shows
the total Fokker-Planck coefficient in comparison with the magnetostatic results (dotted line), the small
scale Fokker-Planck coefficient (dashed line) and the large scale Fokker-Planck coefficient 
(dash-dotted line).}
\label{madisonf2}
\end{figure}

To test the analytical predictions we also calculated the different Fokker-Planck
coefficients numerically. To calculate $D_{\mu \mu}$ or $\lambda_{\parallel}$
as a function of the dimensionless rigidity $R$ we must express the ratio $v_A / v$ through 
the parameter $R$:
\be
{v_A \over v} = {v_A \over c} {\sqrt{R_0^2 + R^2} \over R}
\label{s2e22}
\ee
with
\bdm
R_0 = {k_{min} \over B_0} \cdot
\left\{
\begin{array}{ccc}
0.511 MV & \textnormal{for} & e^{-} \\
938 MV & \textnormal{for} & p^{+}
\end{array}
\right.
\label{s2e23}
\edm
Fig. \ref{madisonf1} and \ref{madisonf2} show the different Fokker-Planck coefficients
for protons and the following set of parameters:
\bdm
s & = & 5/3 \nonumber\\
k_{min} & = & \left( 0.03 AU \right)^{-1} \nonumber\\
\delta B^2 / B_0^2 & = & 0.2 \nonumber\\
B_0 & = & 4.12 \; nT \nonumber\\
R & = & 0.001 \nonumber\\
v_A & = & 33.5 \; km / s\nonumber\\
\alpha & = & 1
\label{s2e24}
\edm
With these parameters we have $R_0 (p^{+}) \approx 0.169$ and $R_0 (e^{-}) \approx 92.1 \cdot 10^{-6}$.
Fig. \ref{madisonf1} shows the numerical calculated Fokker-Planck coefficients for $\epsilon=1$. In this
case we split the integrals at $k=k_{res}$. As expected the small scale and and large scale Fokker-Planck
coefficients are approximatelly equal except for small $\mu$ where the large scale contribution is dominant.
In Fig. \ref{madisonf2} we have shown the numerical results for $\epsilon=1/3$. Here the small scale
contribution is always dominant except for small $\mu$.
\section{Comparison of the small scale and large scale Fokker-Planck coefficients}
In this section we calculate the ratio
\be
\xi:={D_{\mu\mu}^{LS} \over D_{\mu\mu}^{SS}} = \left( {\epsilon \over \mu R} \right)^s {A \over B} - 1
\label{s2e1}
\ee
for the cases $a \mu \ll \epsilon \ll 1$, $\epsilon \ll a \mu \ll 1$ and $\epsilon \ll 1 \ll a \mu$.
\subsection{The case $a \mu \ll \epsilon \ll 1$}
Here we find with Eq. (\ref{s1e19})
\be
\xi (a \mu \ll \epsilon \ll 1) = {\pi s \over 2 \sin (\pi s / 2)} \cdot \left({\epsilon \over a \mu}\right)^s \gg 1
\label{s2e2}
\ee
and therefore
\be
D_{\mu\mu}^{LS} \gg D_{\mu\mu}^{SS}.
\label{s2e3}
\ee
The last both results can also be derived directly from Eq. (\ref{s1e20}).
In the case $a \mu \ll \epsilon \ll 1$ the large scale contribution is much larger then the small scale
contribution.
\subsection{The case $\epsilon \ll a \mu \ll 1$}
In this case we have
\be
\xi (\epsilon \ll a \mu \ll 1) = {2 \sin (\pi s / 2) \over \pi (2-s)} \cdot \left({\epsilon \over a \mu}\right)^{2-s} \ll 1
\label{s2e3a}
\ee
and therefore
\be
D_{\mu\mu}^{LS} \ll D_{\mu\mu}^{SS}.
\label{s2e3b}
\ee
For $\epsilon \ll a \mu \ll 1$ the large scale contribution is much smaller then the small scale
contribution.
\subsection{The case $\epsilon \ll 1 \ll a \mu$}
Here we find with Eq. (\ref{s1e19}) or Eq. (\ref{s1e20})
\be
\xi (\epsilon \ll 1 \ll a \mu) \approx {2 \over \pi (2-s)} \cdot {\epsilon^{2-s} \over a \mu} \ll 1
\label{s2e4}
\ee
and therefore
\be
D_{\mu\mu}^{LS} \ll D_{\mu\mu}^{SS}.
\label{s2e5}
\ee
Also in the case $\epsilon \ll 1 \ll a \mu$ the large scale contribution is much smaller then the small scale
contribution.
\section{The parallel mean free path}
In this section we calculate the parallel mean free path using the results for
the Fokker-Planck coefficients of the last section.
\subsection{The total contribution}
With Eq. (\ref{s1e20}) we have
\bdm
\lambda_{\parallel}^{tot} & = & {3 v \over 4} \int_{0}^{1} d \mu \; {(1-\mu^2)^2 \over D_{\mu\mu}^{tot}(\mu)} \nonumber\\
& \approx & {3 v \over 4} \int_{0}^{1/a} d \mu \; {(1-\mu^2)^2 \over D_{\mu\mu}^{tot}(a \mu \ll 1)} \nonumber\\
& + & {3 v \over 4} \int_{1/a}^{1} d \mu \; {(1-\mu^2)^2 \over D_{\mu\mu}^{tot}(a \mu \gg 1)}
\label{s3e3}
\edm
and we obtain for the parallel mean free path
\bdm
\lambda_{\parallel}^{tot} & \approx & {3 \alpha v_A v k_{min} \over \pi (s-1) \Omega^2} {B_0^2 \over \delta B^2}
{a \over R^s} {2 \over (2-s)(4-s)} \nonumber\\
& \approx &  {6 \over \pi (s-1)(2-s)(4-s)} {B_0^2 \over \delta B^2} \cdot {R^{2-s} \over k_{min}} \nonumber\\
& \approx & \lambda_{\parallel}^{MS}
\label{s3e4}
\edm
which is approximatelly equal to the magnetostatic results ($\lambda_{\parallel}^{MS}$).
For a Kolmogorov spectrum ($s=5/3$) we find that the parallel mean free path is $\sim R^{1/3}$,
whereas for $s=3/2$ the parallel mean free path is $\sim R^{1/2}$.

We also find that the result for $\lambda_{\parallel}^{tot}$ is independent of the damping parameter
$\alpha$. Therefore we come to the conclusion that dynamical effects can be neglected if we
calculate the total parallel mean free path for medium rigidities and for slab geometry.
This result is in agreement with numerical calculations presented in this paper (see Fig. \ref{madisonf1}
and \ref{madisonf2}).
\subsection{The small scale contribution}
If we neglect the large scale contribution if we calculate the parallel mean free path
we have
\bdm
\lambda_{\parallel}^{SS} & = & {3 v \over 4} \int_{0}^{1} d \mu \; {(1-\mu^2)^2 \over D_{\mu\mu}^{SS}(\mu)} \nonumber\\
& \approx & {3 v \over 4} \int_{0}^{1/a} d \mu \; {(1-\mu^2)^2 \over D_{\mu\mu}^{SS}(a \mu \ll \epsilon \ll 1)} \nonumber\\
& + & {3 v \over 4} \int_{1/a}^{1/\epsilon} d \mu \; {(1-\mu^2)^2 \over D_{\mu\mu}^{SS}(\epsilon \ll a \mu \ll 1)} \nonumber\\
& + & {3 v \over 4} \int_{1/\epsilon}^{1} d \mu \; {(1-\mu^2)^2 \over D_{\mu\mu}^{SS}(\epsilon \ll 1 \ll a \mu)}
\label{s3e5}
\edm
and together with Eq. (\ref{s1e20}) we find for the parallel mean free path
\be
\lambda_{\parallel}^{SS} \sim \int_{0}^{1/a} d \mu \; {1-\mu^2 \over \mu^s} + ...
\rightarrow \infty.
\label{s3e6}
\ee
If we neglect the large scale contribution the parallel mean free path goes to infinity. This result
comes from very small values of $\mu$. It should be noted that in this regime QLT is no longer valid because
for $\mu \rightarrow 0$ nonlinear effects are essential (see Shalchi 2005). 
If we would use a more accurate, nonlinear theory instead of QLT, the problem of the infinitely
large parallel mean free path (Eq. (\ref{s3e6})) should vanish. 
\subsection{The large scale contribution}
If we neglect the small scale contribution if we calculate the parallel mean free path
we have
\bdm
\lambda_{\parallel}^{LS} & = & {3 v \over 4} \int_{0}^{1} d \mu \; {(1-\mu^2)^2 \over D_{\mu\mu}^{LS}(\mu)} \nonumber\\
& \approx & {3 v \over 4} \int_{0}^{\epsilon/a} d \mu \; {(1-\mu^2)^2 \over D_{\mu\mu}^{LS}(a \mu \ll \epsilon)} \nonumber\\
& + & {3 v \over 4} \int_{\epsilon/a}^{1} d \mu \; {(1-\mu^2)^2 \over D_{\mu\mu}^{LS}(\epsilon \ll a \mu)}.
\label{s3e7}
\edm
With Eq. (\ref{s1e20}) we obtain for the parallel mean free path
\be
\lambda_{\parallel}^{LS} \approx {3(2-s) \over (s-1)(3-s)(5-s)} {B_0^2 \over \delta B^2} \cdot {R^{2-s} \over k_{min}} {a \over \epsilon^{2-s}}.
\label{s3e8}
\ee
In this case the parallel mean free path is larger than the magnetostatic result. This would be the case when the turbulence is 
damped at a scale larger than the resonant scale.
\section{Discussions and Summary}
In the current paper we calculated the pitch-angle Fokker-Planck coefficient and the parallel
mean free path using QLT. Our intension was to find out how important dynamical effects
are for pure slab geometry and for medium rigidities and whether the small scale
or the large scale contribution controlls the total Fokker-Planck coefficient and the
parallel mean free path.

If we consider Eq. (\ref{s3e4}) we come to the conclusion that we can neclect
dynamical effects if we calculate the parallel mean free path for medium rigidities.
The situation would be different if we would consider small rigidities and if we would use
a power spectrum with dissipation range. This was considered in Teufel \& Schlickeiser 2002 and 2003 for
pure slab geometry and in Shalchi \& Schlickeiser 2003 for pure 2D and composite slab/2D geometry.

In the current paper we compared the different contributions to the Fokker-Planck coefficient.
We find that for $a \mu \ll \epsilon$ the large scale contribution controlls the Fokker-Planck coefficient
whereas for $\epsilon \ll a \mu$ the small scale contribution is dominant.
Thus for slab model turbulence quasilinear approximation is a good approximation except for CRs with
pitch angle close to $90^o$. In this regime ($\mu \approx 0$) nonlinear effects can no longer
be neglected (see Shalchi 2005). Because of the results of the current paper we come
to the conclusion that previous calculations done in the quasilinear limit and for slab geometry are valid.
In non-slab models (e.g. the composite slab/2D model of Bieber et al. 1994) however, nonlinear effects 
always play a significant role, because perpendicular diffusion itself can have a strong influence on 
pitch-angle scattering (see Shalchi et al. 2004).

\section*{Acknowledgments}
This work was supported by the NSF grant AST-0307869 ATM-0312282 and the
NSF Center for Magnetic Self Organization in Laboratory and Astrophysical Plasmas.
This research was also supported by the National Science Foundation
under grant ATM-0000315. 
This work is the result of a collaboration between the University of Bochum, Theoretische Physik IV and
the University of Wisconsin-Madison, Department of Astronomy.

{}

\end{document}